\documentclass[10pt,conference]{IEEEtran}
\usepackage{mathptmx} 

\usepackage{fancyhdr}
\usepackage[normalem]{ulem}
\usepackage[hyphens]{url}
\usepackage[sort,nocompress]{cite}
\usepackage[final]{microtype}
\usepackage[keeplastbox]{flushend}
\usepackage{epsfig,endnotes}
\usepackage{graphicx}
\usepackage{adjustbox}
\usepackage{url}
\usepackage{float}
\usepackage{caption}
\usepackage{subfigure}
\usepackage{stfloats}
\usepackage{enumitem}
\usepackage{hyperref}
\usepackage{pifont}

\usepackage{tikz} 
\usepackage{amsmath}

\usepackage{filecontents}

\usepackage{graphicx,pifont}

\usepackage{bbding}
\usepackage{subfigure}
\usepackage{float}
\graphicspath{{pics/}}

\usepackage{color}

\def\BibTeX{{\rm B\kern-.05em{\sc i\kern-.025em b}\kern-.08em
		T\kern-.1667em\lower.7ex\hbox{E}\kern-.125emX}}

\newcommand{\hpcayear}{2024}

\newcommand{\hpcasubmissionnumber}{512}
\title{LearnedFTL: A Learning-based Page-level FTL for Reducing Double Reads in Flash-based SSDs}

\def\hpcacameraready{} 

\newcommand\hpcaauthors{Shengzhe Wang$^\dagger$, Zihang Lin$^\dagger$, Suzhen Wu$^\dagger$, Hong Jiang$^\ddagger$, Jie Zhang$^\ast$ and Bo Mao$^\dagger$}
\newcommand\hpcaaffiliation{$^\dagger$Xiamen Key Laboratory of Intelligent Storage and Computing, School of Informatics, \\Xiamen University, Xiamen, Fujian, China\\
$^\ddagger$Computer Science and Engineering Department, University of Texas at Arlington, Arlington, TX, USA\\
$^\ast$School of Computer Science,
Peking University, Beijing, China}
\newcommand\hpcaemail{Corresponding Author: Suzhen Wu (suzhen@xmu.edu.cn) and Bo Mao (maobo@xmu.edu.cn)}

\author{
  \ifdefined\hpcacameraready
    \IEEEauthorblockN{\hpcaauthors{}}
      \IEEEauthorblockA{
        \hpcaaffiliation{} \\
        \hpcaemail{}
      }
  \else
    \IEEEauthorblockN{\normalsize{HPCA \hpcayear{} Submission
      \textbf{\#\hpcasubmissionnumber{}}} \\
      \IEEEauthorblockA{
        Confidential Draft \\
        Do NOT Distribute!!
      }
    }
  \fi 
}

\fancypagestyle{camerareadyfirstpage}{%
  \fancyhead{}
  
  \fancyhead[C]{
    \ifdefined\aeopen
    \parbox[][12mm][t]{13.5cm}{\hpcayear{} IEEE International Symposium on High-Performance Computer Architecture (HPCA)}    
    \else
      \ifdefined\aereviewed
      \parbox[][12mm][t]{13.5cm}{\hpcayear{} IEEE International Symposium on High-Performance Computer Architecture (HPCA)}
      \else
      \ifdefined\aereproduced
      \parbox[][12mm][t]{13.5cm}{\hpcayear{} IEEE International Symposium on High-Performance Computer Architecture (HPCA)}
      \else
      \parbox[][0mm][t]{13.5cm}{\hpcayear{} IEEE International Symposium on High-Performance Computer Architecture (HPCA)}
    \fi 
    \fi 
    \fi 
    \ifdefined\aeopen 
      \includegraphics[width=12mm,height=12mm]{ae-badges/open-research-objects.pdf}
    \fi 
    \ifdefined\aereviewed
      \includegraphics[width=12mm,height=12mm]{ae-badges/research-objects-reviewed.pdf}
    \fi 
    \ifdefined\aereproduced
      \includegraphics[width=12mm,height=12mm]{ae-badges/results-reproduced.pdf}
    \fi
  }
  \fancyfoot[C]{}
}
\begin{document}
\maketitle

\ifdefined\hpcacameraready 
  \thispagestyle{camerareadyfirstpage}
  \pagestyle{empty}
\else
  \thispagestyle{plain}
  \pagestyle{plain}
\fi

\newcommand{\hpcaheight}{0mm}
\ifdefined\eaopen
\renewcommand{\hpcaheight}{12mm}
\fi

\begin{abstract}
We present LearnedFTL, a new on-demand page-level flash translation layer (FTL) design, which employs learned indexes to improve the address translation efficiency of flash-based SSDs. The first of its kind, it reduces the number of double reads induced by address translation in random read accesses. LearnedFTL proposes three key techniques: an in-place-update linear model to build learned indexes efficiently, a virtual PPN representation to obtain contiguous PPNs for sorted LPNs, and a group-based allocation and model training via GC/rewrite strategy to reduce the training overhead. By tightly integrating the aforementioned key techniques, LearnedFTL considerably speeds up address translation while reducing the number of flash read accesses caused by the address translation. Our extensive experiments on a FEMU-based prototype show that LearnedFTL can reduce up to 55.5\% address translation-induced double reads. As a result, LearnedFTL reduces the P99 tail latency by 2.9$\times$ $\sim$ 12.2$\times$ with an average of 5.5$\times$ and 8.2$\times$ compared to the state-of-the-art TPFTL and LeaFTL schemes, respectively. 
 
\end{abstract}

\section{Introduction}
With the emergence of 3D NAND and NVMe techniques~\cite{GSSA-HPCA21,MGC-HPCA23,AstriFlash-HPCA23}, the capacity and performance of flash-based SSDs have increased significantly. Accordingly, the mapping table size for those using page-level Flash Translation Layer (FTL) cannot be practically deployed in SSDs since it consumes a prohibitively large memory. Moreover, storing the mapping table in SSDs causes the double-read problem of constantly accessing the flash for the address translation, which directly degrades the read performance of SSDs.

To solve this problem, the mainstream demand-based FTL (DFTL) proposes a mapping cache method \cite{gupta2009dftl,zhou2015efficient,chae2020dsftl}. Specifically, to reduce the memory overhead of the page-level mapping table, DFTL exploits the locality of workloads and sets a small-sized mapping cache in SSD internal memory to cache frequently accessed mappings. If an LPA cannot be found in this mapping cache, DFTL needs an extra flash read to fetch the corresponding PPA, followed by another flash read to access the data. This phenomenon is referred to as \emph{double reads}. Since the mapping cache can capture locality well, DFTL generates only a small number of double reads under workloads with high locality, which in turn has minor impacts on SSD performance.

However, the double-read problem in DFTL becomes very severe under random reads. Since there is no obvious data locality in random reads, and the LPNs of continuous requests may separate far away, it is difficult for the mapping cache to buffer the mappings in need. As a result, almost all read requests require double reads~\cite{gupta2009dftl,zhou2015efficient}, which incurs poor random read performance as also validated in the subsection~\ref{two-one}. Unfortunately, in many SSD-based modern applications, the proportion of random access is gradually increasing and even occupies a dominant position. The ability to handle random access becomes critical for FTL.

How to index the most mappings in cache-sized DRAM memory is the key to increasing the hit rate of random reads. Recent studies on learned index~\cite{ALEX,APEX,kraska2018case,wei2020fast} have demonstrated the feasibility of realizing this goal. Learned index builds machine learning models based on the key-position mappings. With these models, the learned index can index hundreds of mappings with a few parameters. Ideally, by adopting the learned index to all LPN-PPN mappings, one can calculate the PPN of an LPN directly from learned models without double reads.

Unfortunately, the limitations of the learned index pose several challenges. First, the accuracy of the learned index cannot reach 100\%. When the PPN prediction is incorrect, the neighboring flash pages to the predicted PPN need to be probed in order to identify the correct PPN. However, this process results in degraded performance. Second, learned indexes require assigning consecutive PPNs to consecutive LPNs, which conflicts with the internal access parallelism of SSDs. Third, the complex model training of learned indexes will cause additional performance and space overheads. Specifically, training a learned index needs multiple time-consuming computational operations on the critical write path. Moreover, training random write requests requires more space to store the learning models.

\begin{figure*}[t]
    \centering
    \includegraphics[scale=0.75]{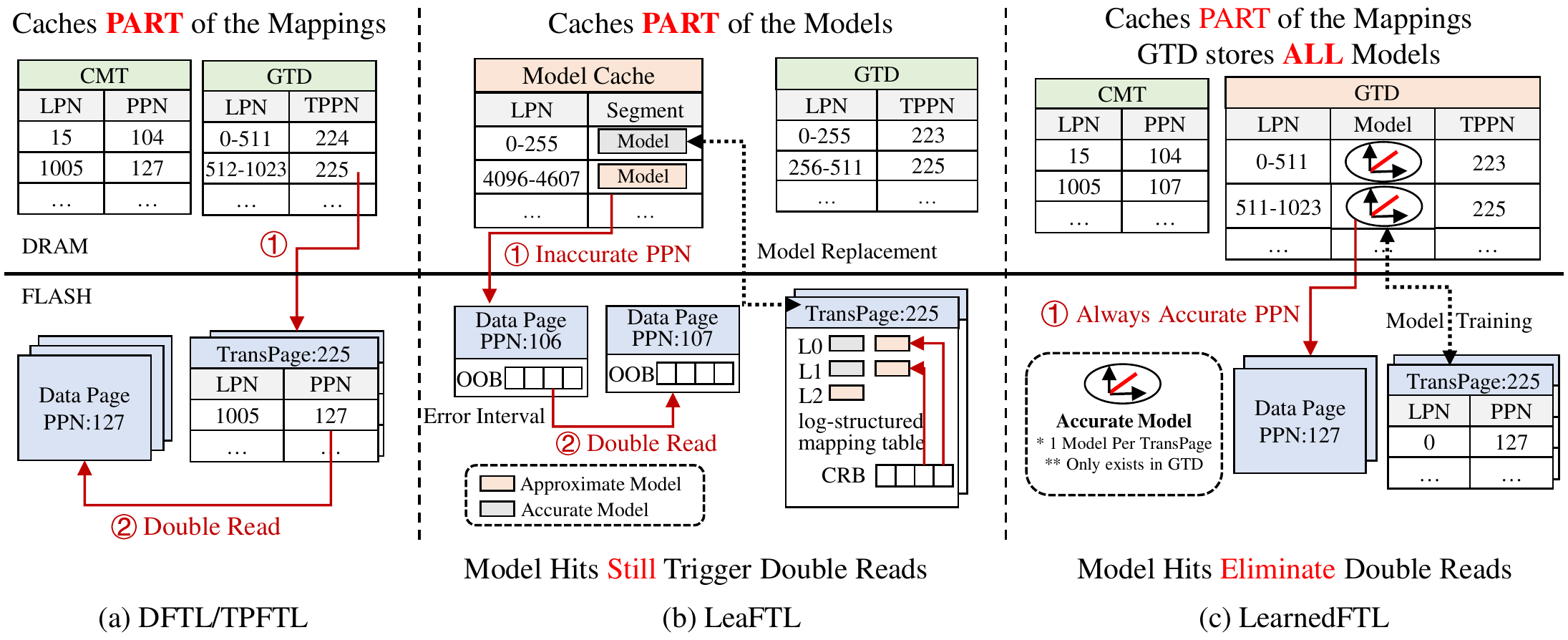}

    \caption{Representative mapping schemes in flash-based SSDs.}
    \label{three_schemes}
\end{figure*}

To tackle these challenges, we propose LearnedFTL, a learning-based page-level FTL. LearnedFTL combines the learned indexes with the existing demand-based page-level FTL scheme, TPFTL~\cite{zhou2015efficient}, where TPFTL handles the locality workloads, and learned indexes handle the random workloads. Each learned index in LearnedFTL is a piece-wise linear model with adjustable parameters, called the in-place-update linear model. To ensure the accuracy of the model prediction, LearnedFTL equips each in-place-update linear model with a bitmap filter where each bit indicates whether the prediction of an LPN is accurate. To obtain contiguous PPNs for sorted LPNs with SSD's internal parallelism, LearnedFTL proposes a virtual PPN representation to convert the incontiguous PPNs to sequential ones. To reduce the space overhead of model training under random writes, a group-based allocation strategy is proposed to replace the current dynamic allocation strategy. Lastly, LearnedFTL proposes two model training strategies to minimize the performance overhead: first, initializing the model based on sequential write requests, and second, training the model through garbage collection. Overall, this paper makes the following contributions:

\begin{itemize}
\item Our experimental analysis of the random reads in flash-based SSDs reveals that the address-translation-induced double-read accesses to flash storage are the root cause of SSD’s poor random-read performance.

\item We propose a novel FTL design, LearnedFTL, to improve the read performance of SSDs.
The innovation of LearnedFTL lies in its ability to effectively combine Learned Index and address mapping in FTL that maximizes performance while minimizing modifications. By incorporating learned indexes into the currently popular demand-based FTL, LearnedFTL can enhance random read performance without compromising the ability to handle workloads with locality.
To the best of our knowledge, LearnedFTL is the first FTL design that optimizes for random read performance.
    
\item We propose several optimizations in LearnedFTL to facilitate the learned indexes, including an in-place-update linear model equipped with a bitmap filter to guarantee the accuracy of predictions, a virtual PPN representation to convert the unordered PPNs from different parallel units into contiguous ones, a group-based allocation strategy and two model training strategies to reduce training overheads.
    
\item We implement the prototype of LearnedFTL on the SSD emulator FEMU~\cite{li2018case}. The extensive evaluations validate the efficacy of LearnedFTL over the state-of-the-art FTLs.
\end{itemize}

\section{Background and Motivation}
\label{background}

\subsection{Demand-based Page-Level FTLs}
\label{two-one}

Address translation is a vital function in FTL, which searches the physical addresses of flash memory for incoming requests. There exist several mapping schemes such as page-level mapping, block-level mapping, and hybrid mapping~\cite{lee2008last,lee2006fast,bez2003introduction}.
Since the flash page is the basic unit for read/write operations, page-level mapping can handle requests flexibly and performs well. However, this fine-grained mapping scheme requires a huge DRAM memory to accommodate its mapping table. For example, suppose an SSD has a 10TB capacity with a 4KB page size, and each entry of the LPN-PPN mapping is 8B, the SSD requires 20GB of DRAM memory to store the LPN-PPN mapping of 2.5 billion entries. This huge DRAM memory consumption, unfortunately, is impractical for enterprise SSDs.

Block-level mapping~\cite{kang2006superblock} and hybrid mapping~\cite{lee2006fast,lee2008last} addressed the space issue by compressing the mapping table. In these schemes, the address mappings are organized at the granularity of a flash block, leading to a significantly lower mapping space overhead. However, block-level mapping has a limitation, that is, data stored in a block must have contiguous LPNs. Flash pages can only be written to a fixed location in the flash block. Consequently, these mapping schemes exhibit poor writing performance.

To strike a good balance between write performance and DRAM memory, demand-based page-level FTLs (DFTL)~\cite{gupta2009dftl} is proposed. Specifically, DFTL uses a selective cache solution to only buffer frequently accessed mappings into SSD memory to exploit workloads' temporal locality, thus reducing memory usage without compromising performance. Figure~\ref{three_schemes}(a) illustrates the general structure of DFTL. It stores the whole mapping table in multiple flash pages, called \textbf{translation pages}. DFTL contains two data structures in SSD memory. \textbf{Cached Mapping Table (CMT)} stores mapping information for frequently accessed flash pages while \textbf{Global Translation Directory (GTD)} records the physical location of translation pages in flash memory. For requests that miss from CMT, DFTL imposes a high miss penalty. In particular, the SSD controller needs an additional flash read to fetch the missing mapping from the translation page. A read request may generate two flash reads for data and metadata, which is called \textbf{double reads}.

Several demand-based page-level FTLs have been proposed to address double reads by exploiting workload locality characteristics. Examples include TPFTL~\cite{zhou2015efficient}, HCFTL~\cite{chen2019hcftl}, and ZFTL~\cite{wang2012zftl}. Among them, TPFTL is a well-known FTL that utilizes both temporal and spatial locality. It proposes a workload-adaptive loading policy to prefetch mappings to CMT based on the request length. This approach improves the hit ratio of CMT and significantly alleviates double reads in workloads with a strong locality.

\subsection{Performance Impact of Double Reads}
Despite the effectiveness of demand-based page-level FTLs in reducing memory usage, their efficiency is limited to workloads with high locality. This becomes problematic in some modern applications with random accesses. Therefore, the ability of FTL to solve double reads under random workloads is of great importance. 

\begin{figure}[t]
    \subfigure[Read throughput]{
        \begin{minipage}[t]{0.46\linewidth}
            \centering
            \includegraphics[scale=0.82]{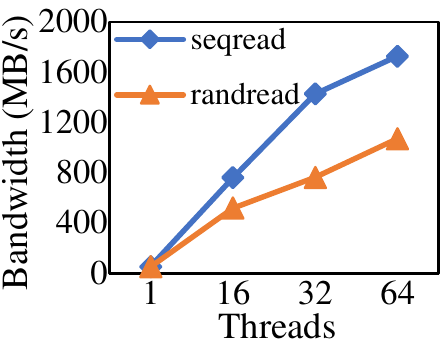}
            \label{seqrand}
        \end{minipage}%
    }
    \subfigure[CMT hit ratio]{
        \begin{minipage}[t]{0.46\linewidth}
            \centering
            \includegraphics[scale=0.80]{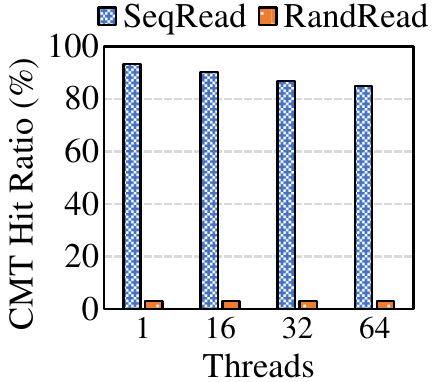}
            \label{breakdown}
        \end{minipage}
    }
    \caption{The performance of a FEMU-emulated SSD under sequential reads and random reads.} 
    \label{first-test}
\end{figure}

To investigate whether demand-based FTLs can handle random workloads, we evaluate the sequential and random read performance of TPFTL driven by FIO~\cite{FIOmisc} stress testing tool. As shown in Figure~\ref{seqrand}, regardless of the variation in the number of threads, the performance of random reads consistently falls short compared to sequential reads (i.e., up to 60\% degradation). Figure~\ref{breakdown} shows the CMT hit ratio under different threads. Under random reads, although TPFTL adopts a prefetching strategy, it can only predictively prefetch PPNs near one PPN. However, the two following requests in random reads may be far apart. As a result, the prefetching strategy becomes ineffective, incurring a very low CMT hit ratio.

\begin{figure}[t]
    \centering
    \includegraphics[scale=0.84]{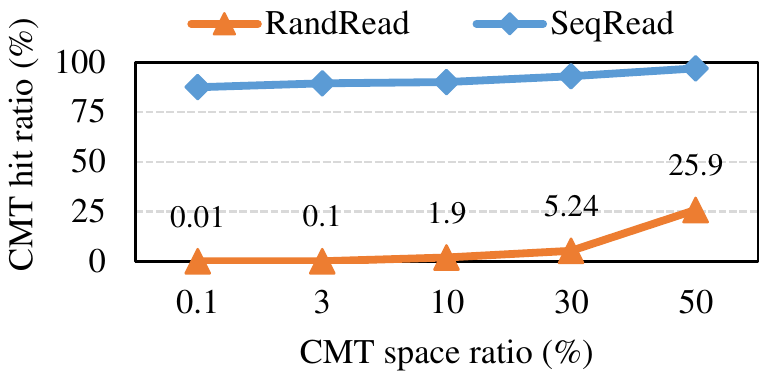}  
    \caption{The hit ratio of TPFTL under different CMT space.}
    \label{diff_cmt_space}
\end{figure}

Increasing the size of the CMT is a straightforward solution to improve the random read performance. However, this approach remains ineffective due to cache contention. Figure~\ref{diff_cmt_space} illustrates the changes in the CMT hit ratio of TPFTL when increasing CMT space. Even when the CMT space expands to 50\% of the total page mappings, the hit ratio only improves slightly to 25.9\%. It is clear that contention for the CMT will exist unless the CMT can accommodate the majority of the mappings. Consequently, regardless of the practical capacity of the CMT, the prefetched mappings will be frequently replaced, leading to a low CMT hit ratio. 

The above experiments and analysis demonstrate that the selective cache solution of demand-based FTL cannot handle random reads. Since a random read request may access any LPN in the entire address space, an efficient solution is to place as many mappings as possible in the small capacity of SSD memory. The mapping table compression scheme mentioned earlier is the only approach that meets these criteria. However, this approach adversely impacts SSD write performance. Thus, to improve the random read performance, we need a new solution to compress the mapping table without degrading SSD write performance.

\subsection{Learned Index and LeaFTL}
\label{two-four}
\begin{figure}[t]
    \centering
    \includegraphics[scale=0.9]{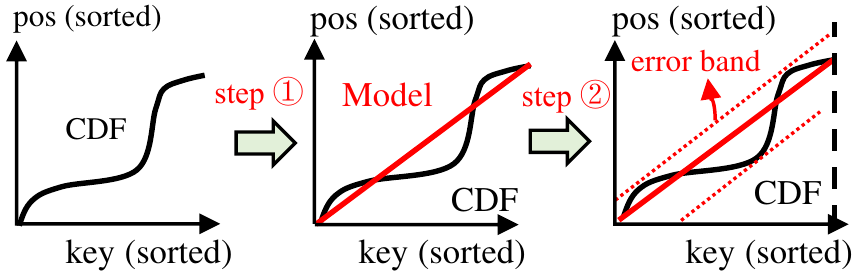}
    \caption{The workflow of the learned index.}
    \label{fig-principle}
\end{figure}

\noindent \textbf{Learned Index.}
Recent studies on \emph{learned index}~\cite{kraska2018case,ALEX,ferragina2020pgm,li2021finedex,Tsunami,ROLEX} have demonstrated its potential for compressing the mapping table since it has a high compression rate and low write limit. The learned index only requires the LPN-PPN mappings to be ordered and builds lightweight models for key-position mappings. A model with several parameters can calculate hundreds of data locations, thus reducing memory consumption.
Figure~\ref{fig-principle} illustrates the workflow of the learned index. Building a learned index model only requires two steps: training an approximate model (usually linear models) over the key-position mappings and identifying the maximum error between the fitted model and the actual values. With the simple model and maximum error, the needed value can be found in the error interval [$y-error$, $y+error$], where $y$ is the predicted position by the approximate model. 

\noindent \textbf{LeaFTL~\cite{sun2023leaftl}.}
Ideally, if the learned index could completely replace the existing mapping table structure and index all mappings in memory, the double-read problem could be solved. Recently, LeaFTL has taken this approach. The primary motivation behind LeaFTL is to utilize the learned indexes to replace the current mapping table, thereby reducing the DRAM memory overhead to store the mapping table.

Figure~\ref{three_schemes}(b) illustrates the structure of LeaFTL. LeaFTL uses a learned segment design for learned indexes, and each learned segment has four parameters \emph{\textbf{[S, K, L, I]}}, expressed as a model $PPN = LPN * K + I, LPN\in[S, S+L]$. In LeaFTL's configuration, one learned segment can index up to 256 mappings. For learned segments that are not 100\% accurate (denoted as approximate segments), LeaFTL conceals the error interval in the Out Of Band (OOB) area of each flash page. When the model predicts a wrong PPN, LeaFTL reads the error interval from the OOB of the mispredicted flash page and finds the correct PPN, then LeaFTL can read the correct PPN to access data. With this approach, each misprediction requires 2 flash reads.

Since the learned index cannot be updated unless retrained, LeaFTL adopts the idea of a Log-Structured Merge-tree to ensure the timeliness of the learned segments. LeaFTL allocates a small area in SSD internal memory, called \textbf{data buffer}, to buffer newly written data (up to 2048 pages). When the data buffer is full, LeaFTL sorts all data by their LPNs and then writes them to flash pages of continuous PPNs. After that, LeaFTL groups these mappings according to the translation pages they belong, and each group trains a newly learned segment. Then all the learned segments are flushed to the corresponding translation pages. In each translation page, the learned segments are organized in a log-structured mapping table (LSMT). The newly created segment is inserted into the top layer. If one layer has overlapped segment, LeaFTL will migrate the old segment to the next layer. 
Since the log-structured design brings space amplification (In our evaluations, LSMT can only reduce the space to 10\%-15\% of the original mapping table, which is still too large to be fully stored in memory), LeaFTL continues to use the idea of CMT and only caches the most frequently used learned segments into memory. 

\subsection{Challenges in Learned Indexes/LeaFTL}
While LeaFTL can reduce the size of the mapping table by several times, it, unfortunately, fails in improving the read performance. We take an in-depth analysis and observe multiple key challenges in Learned Index/LeaFTL.

\begin{figure}[t]
    \centering
    \includegraphics[scale=0.9]{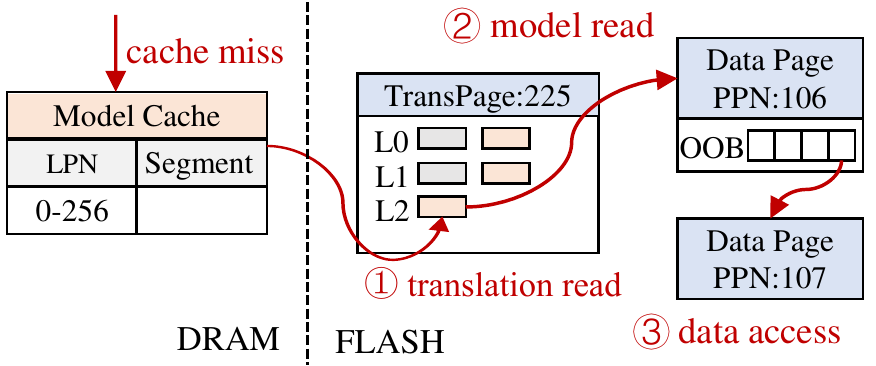}
    \caption{The workflow of triple reads in LeaFTL~\cite{sun2023leaftl}.}
    \label{fig-ptriple}
\end{figure}

\noindent \textbf{Challenge \#1: Accuracy of learned indexes.} The accuracy of learned indexes directly determines the efficiency of address translation. LeaFTL is a purely learned index based address translation scheme and replaces the mapping cache of DFTL/TPFTL with a model cache. Thus, mispredictions of learned indexes will bring \textbf{double reads} (one for error interval in OOB and one for data) in LeaFTL. LeaFTL uses a linear regression model~\cite{kraska2018case,li2021finedex,APEX} that can only express \emph{PPN=LPN*K+B}. As a result, if the LPN-PPNs in the model buffer are not linear, part of the requests may experience double reads.

Moreover, LeaFTL even causes \textbf{triple reads} owing to its model cache design. Figure~\ref{fig-ptriple} illustrates the workflow of triple reads in LeaFTL. When an LPN fails to hit any model in the model cache, it initiates a translation read to find the corresponding model from NAND flash. However, as the model in LeaFTL is an approximate one, the predicted PPN may be wrong. After sending a second flash read to access the wrong flash page, this request has to find the correct PPN via the error interval stored in OOB. Finally, this request reads the correct PPN to access data with a third flash read. The workflow of triple reads indicates that the miss penalty in LeaFTL is much higher than double reads in DFTL.

\begin{figure}[t]
    \subfigure[Random read]{
        \begin{minipage}[t]{0.46\linewidth}
            \centering
            \includegraphics[scale=0.75]{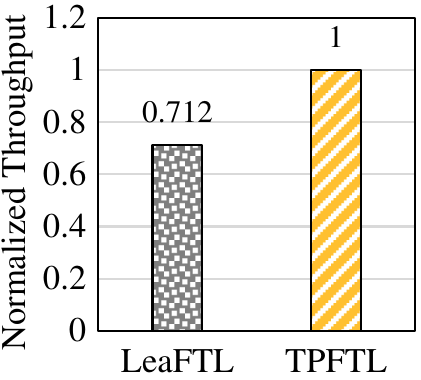}
            \label{leaf_rand}
        \end{minipage}%
    }
    \subfigure[Multi-read count statistics]{
        \begin{minipage}[t]{0.46\linewidth}
            \centering
            \includegraphics[scale=0.7]{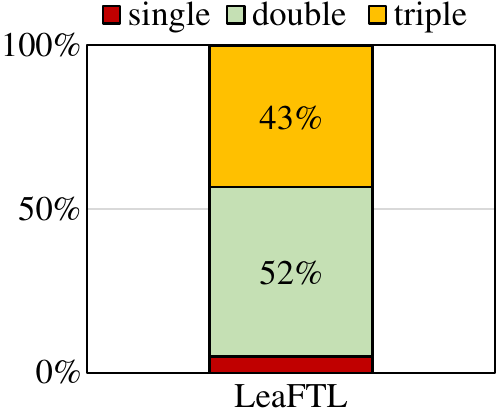}
            \label{stat}
        \end{minipage}
    }    
    \caption{The performance results under random reads.}    
    \label{second-test}
\end{figure}

Considering the fact that the accuracy of learned indexes cannot reach 100\%, the problems of double reads and triple reads will have a significant impact on the performance of LeaFTL. Figure~\ref{leaf_rand} illustrates the normalized throughput of TPFTL and LeaFTL under FIO~\cite{FIOmisc} random reads. LeaFTL exhibits a 29\% lower throughput compared to TPFTL. Figure~\ref{stat} shows the fraction of double reads and triple reads during random reads. Triple reads and double reads account for 43\% and 52\%, respectively. These results demonstrate that the double reads and triple reads make LeaFTL completely unable to handle random reads.

\begin{figure}[t]
    \subfigure[Throughput]{
        \begin{minipage}[t]{0.55\linewidth}
            \centering
            \includegraphics[scale=0.8]{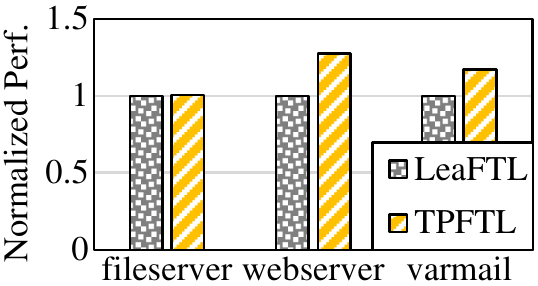}
            \label{leaf_filebench}
        \end{minipage}%
    }
    \subfigure[Hit ratio in webserver]{
        \begin{minipage}[t]{0.38\linewidth}
            \centering
            \includegraphics[scale=0.8]{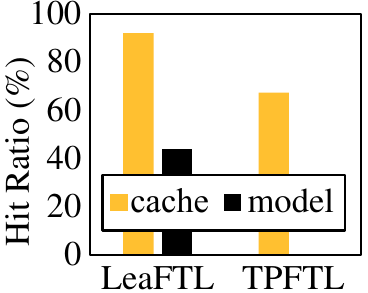}
            \label{hit_ratio_filebench}
        \end{minipage}
    }    
    \caption{The performance of TPFTL and LeaFTL under workloads with high locality.}    
    \label{first-filebench-test}
\end{figure}

Besides affecting random workload, double and triple reads also have a negative effect on workloads with high locality. Figure~\ref{leaf_filebench} shows the performance comparison between TPFTL and LeaFTL under three Filebench~\cite{Filebench} workloads. The performance of LeaFTL is equal to or even worse than that of TPFTL. Figure~\ref{hit_ratio_filebench} shows the cache and model hit ratio under \textbf{webserver} workload (read-intensive). In this context, the cache hit ratio of LeaFTL simply indicates that the model cache contains the corresponding model for the queried LPN, and it does not mean that the correct PPN has been calculated. Due to the space efficiency of learned indexes, the corresponding model of an LPN can be easily found in the model cache, making the model cache hit ratio high. However, there are instances where models experience mispredictions, leading to a significant number of requests requiring double reads. Consequently, the proportion of LPNs that are successfully hit in the model cache and accurately predicted by the model is significantly lower than the LPNs hit in the CMT of TPFTL.
Therefore, TPFTL performs much better than LeaFTL under workloads with high locality. This experiment indicates that when dealing with locality-based workloads, using direct mapping in the cache is more reliable and efficient than using models.

The above analysis and experiments all indicate that handling mispredictions caused by incompletely accurate learned index models directly affects SSD performance.

\noindent \textbf{Challenge \#2: Conflict between the linear model and access parallelism.} 
A key requirement for training the linear model in learned indexes is sorted LPN-PPN mappings. In LeaFTL, after the model buffer is sorted with LPNs, LeaFTL needs to allocate contiguous PPNs for these LPNs. However, modern SSDs are highly dependent on internal parallelism so that multiple flash blocks can be accessed simultaneously across separate flash chips~\cite{hu2011performance,ParaFS}. To be specific, when a set of LPNs needs to be written to an SSD, these LPNs are written to different parallel units (channels, chips, dies, and planes). Since the parallel units belong to a high hierarchical structure, the PPNs in different parallel units may be far apart. Therefore, assigning contiguous PPNs for sorted LPNs is hard in the parallel writing strategy.

\noindent \textbf{Challenge \#3: High training overhead.} 
In LeaFTL, model training is performed on the critical write path. It brings two overheads: (1) Performance overhead: The model training includes sorting, parameters fitting, and compaction, which is time-consuming. These operations performed on the critical write path will directly affect write performance. (2) Space overhead: The space overhead happens in random writes. The LPNs of adjacent write requests are dramatically separated. It is difficult for the model buffer to gather these LPNs in LeaFTL. In the worst case, each LPN-PPN mapping in the model buffer becomes an individual learned segment, leading to a huge space overhead.

To sum up, recent advances in the learned index have shown that it can achieve significantly faster lookup speed and index space savings. Motivated by the urgent need to resolve the double-read problem caused by random reads in flash-based SSDs, along with the challenges learned from learned indexes/LeaFTL, we propose LearnedFTL, which utilizes lightweight learned index models in the existing on-demand page-level FTL (TPFTL) to enhance the random-read performance of flash-based SSDs. 
\section{Design}\label{design}
\subsection{LearnedFTL Overview}

The main idea of LearnedFTL is combining the learned index with demand-based FTL, where the demand-based mapping scheme handles locality-based access patterns and learned indexes handle random access patterns. This design allows LearnedFTL to serve all types of workloads efficiently. Figure~\ref{three_schemes}(c) illustrates the system overview of LearnedFTL.

In LearnedFTL, each request first checks the CMT. If the CMT fails, LearnedFTL queries the corresponding GTD entry and uses the learned index model to predict the PPN. If the prediction is correct, LearnedFTL accesses the predicted PPN directly, thus eliminating the flash double-read operation. If the prediction is inaccurate, LearnedFTL accesses the data by using the original flash double-read method in TPFTL.

Each model in the GTD is called an \textbf{in-place-update linear model}. Each in-place-update model is a piece-wise linear model, and each linear model has adjustable parameters. To guarantee the accuracy of the model predictions (\textbf{Challenge \#1}), each in-place-update linear model is equipped with a bitmap filter, which indicates whether the prediction of a certain LPN is accurate, thus reducing the cost of inaccurate predictions. To obtain contiguous PPNs for sorted LPNs (\textbf{Challenge \#2}), LearnedFTL proposes the virtual PPN (VPPN) representation to convert PPNs from different parallel units into sequential ones. To reduce the space overhead and performance overhead of model training under random writes (\textbf{Challenge \#3}), LearnedFTL proposes a group-based allocation to bring LPNs belonging to the same GTD entry together and proposes two model training strategies, including a computation-free sequential initialization and a model training via GC/rewrite strategy.

\begin{figure}[t]
    \centering
    \includegraphics[scale=0.62]{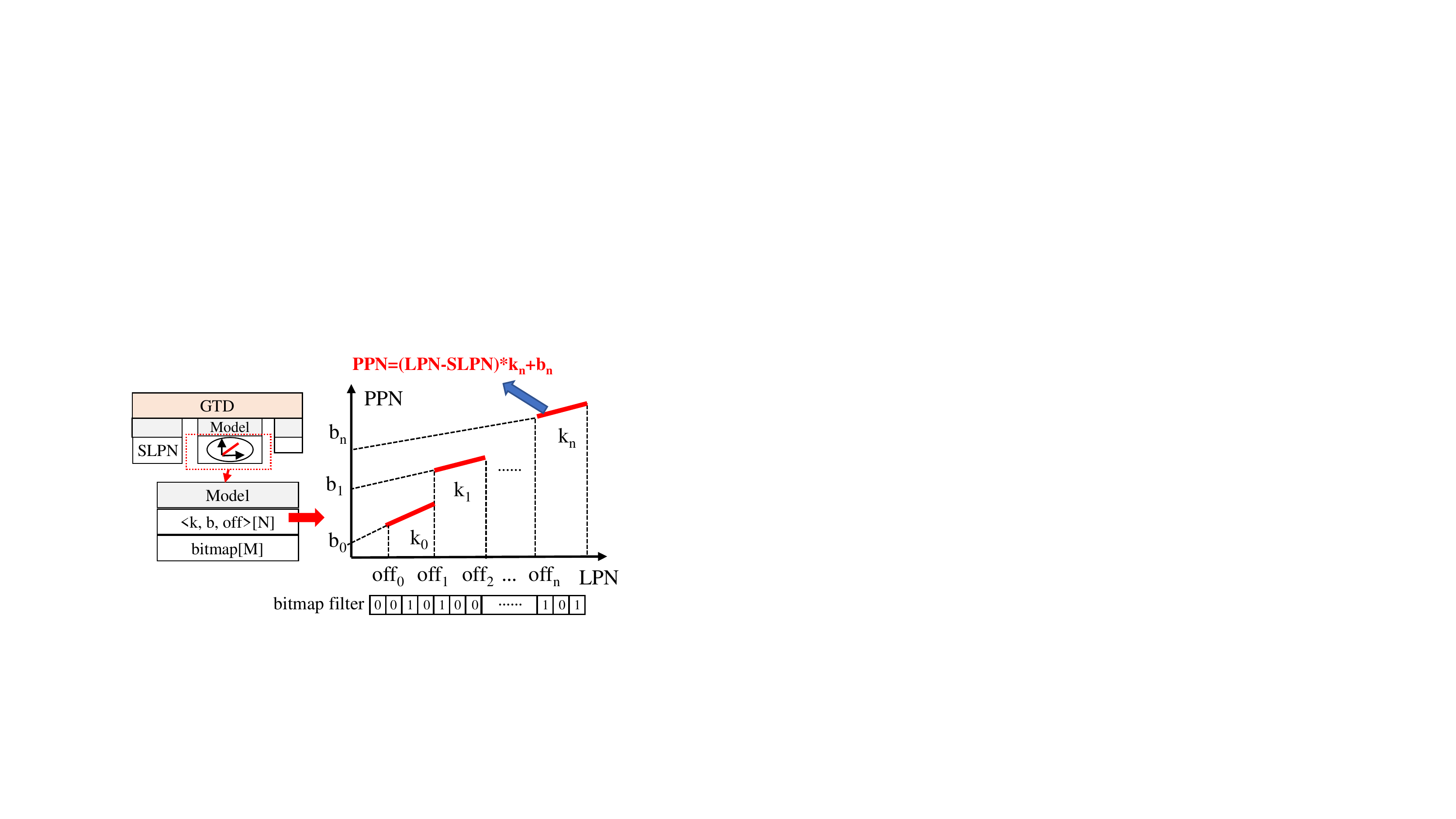}
    \caption{The structure of an in-place-update model in GTD.}
    \label{fig-model-layer}
\end{figure}

\subsection{In-Place-Update Linear Model}
\label{in-place-model}

The model layer in GTD is the most critical component in LearnedFTL, as it determines the efficiency of the entire address mapping process. Figure~\ref{fig-model-layer} illustrates the structure of the \textbf{in-place-update linear model} used in LearnedFTL. Since each model is attached to a GTD entry, each model is only used to predict the mappings of the LPN range represented by its attached GTD entry. An in-place update linear model is a piece-wise linear regression model (PLR model), and it consists of two parts: a \textbf{parameter array} $<$k,b, off$>$[N] and a \textbf{bitmap filter}.

In the parameter array, 
Each $<$k,b,off$>$ represents a linear model, including intercept (\textbf{$\mathrm{b}$}), slope (\textbf{$\mathrm{k}$}), and the offset (\textbf{$\mathrm{off}$}) from this PLR model's starting LPN. Given a certain LPN, the offset (\textbf{$\mathrm{off_x}$}) from the starting LPN is calculated first, and then LearnedFTL queries the corresponding linear model \textbf{$\mathrm{<k_n, b_n, off_n>}$} based on the \textbf{$\mathrm{off_x}$}. The PPN can be predicted using the $y=k_n\times(LPN-LPN_{start}) + b_n$.

\begin{figure}[t]
    \centering
    \includegraphics[scale=0.85]{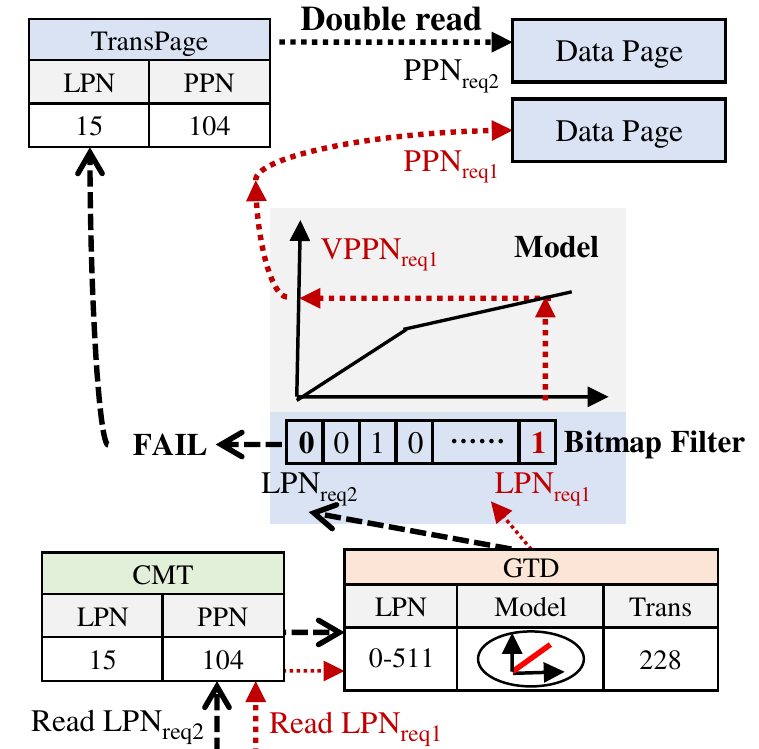}
    \caption{The workflow of bitmap filter.}
    \label{fig-model-predict}
\end{figure}

A bitmap filter is a bitmap, and each bit in the bitmap is associated with an LPN, representing whether an LPN can be accurately predicted (\emph{1} means accurate, \emph{0} means inaccurate). The bitmap is updated during model training (detailed in Section~\ref{three-three}).
With the bitmap filter, the in-place-update linear model offers two significant benefits over the traditional learned indexes: 

(1) \textbf{Accurate predictions}. The bitmap filter can mark which LPNs can make accurate predictions, assisting models to make only accurate predictions.
Figure~\ref{fig-model-predict} illustrates the two different instances of the bitmap filter. For a request with an $LPN_{req1}$ that needs to use the model to predict the PPN, LearnedFTL first checks the corresponding bit in the bitmap and finds the bit is \emph{1}. Then LearnedFTL will perform model prediction to generate the true $PPN_{req1}$. Since this prediction is marked as accurate, LearnedFTL directly uses this $PPN_{req1}$ to access data. For another request with $LPN_{req2}$ whose corresponding bit is \emph{0}, LearnedFTL will perform a double read for this LPN and not use the model to make predictions. With the bitmap filter, LearnedFTL can make only the correct model predictions and avoid miss penalty caused by wrong model predictions. 

\begin{figure}
   \centering
   \setlength{\abovecaptionskip}{5pt}
   \includegraphics[scale=0.65]{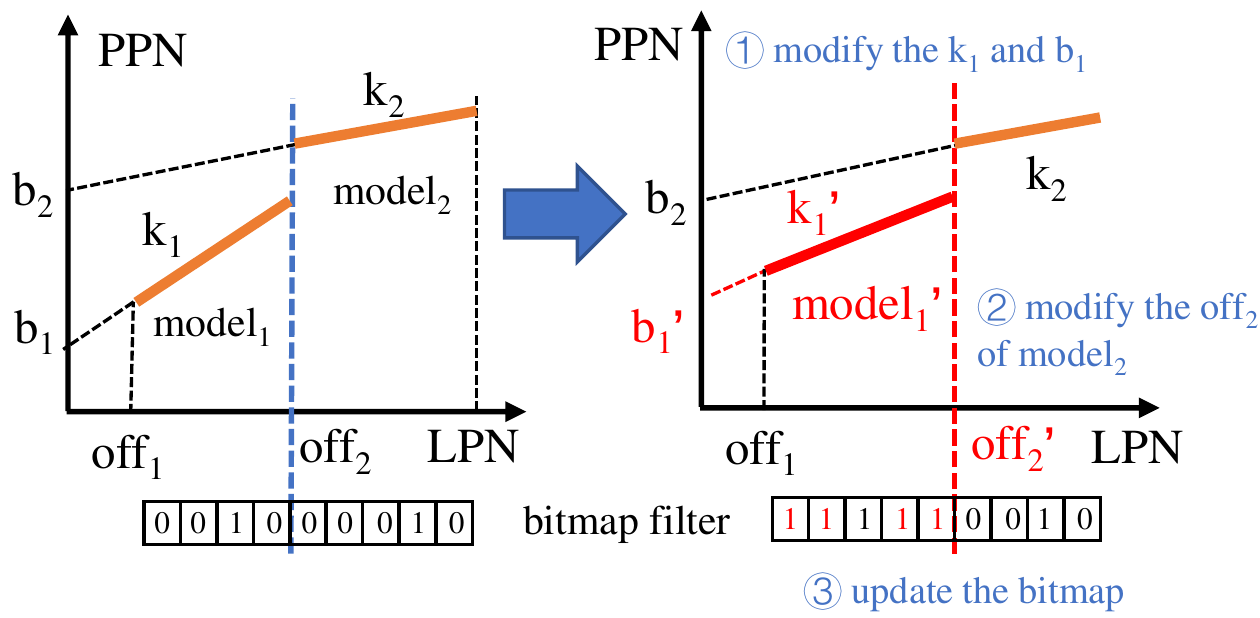}
   \caption{The workflow of model in-place update.}
   \label{fig-inplace-model}\vspace{-10pt}
\end{figure}

(2) \textbf{The model parameters can be updated as needed}. The bitmap filter offers the ability to control each LPN, making in-place update of the model possible. Figure~\ref{fig-inplace-model} shows the workflow of model in-place update, when we retrain the LPN-PPN mapping for $model_1$ (with $k_1$ and $b_1$), we can directly update the original model in-place. The model in-place update first modifies the slope $k_1$ and then intercepts $b_1$ to the newly calculated value $k_1'$ and $b_1'$. Since the range of new $model_1'$ and the range of $model_2$ have conflict, the $off_2$ of $model_2$ should be increased until it does not conflict with the new $model_1'$. Finally, the bitmap is updated based on the accuracy of the newly generated model. With the in-place-update ability, an in-place update linear model can always maintain a fixed space overhead, avoiding the need for space compaction like the LSMT in LeaFTL.

The data consistency of the in-place-update linear model is guaranteed upon each update. Specifically, for each write request with an LPN, LearnedFTL first checks if the corresponding bit of this LPN in the bitmap is \emph{\textbf{1}}. If so, LearnedFTL will set this bit to \emph{\textbf{0}} to prevent the model from making wrong predictions.

Since persisting the models to flash upon each update will bring additional writing overhead,
the models are saved to flash follows the GTD saving procedure as the TPFTL and DFTL handle. During a normal power-off, the models are saved in a flash area alongside GTD. This allows us to easily retrieve and use the stored models when the device reboots.
In the event of a power failure, GTD is rebuilt by scanning all translation pages. Models can also be reconstructed from the mapping information within these translation pages, similar to TPFTL and DFTL. The reconstruction won't take much time since the time overhead for model training is minimal, as shown in Figure~\ref{fig-compute-simulation}.

\subsection{Virtual PPN Representation}

\begin{figure}[t]
    \centering
    \includegraphics[scale=0.85]{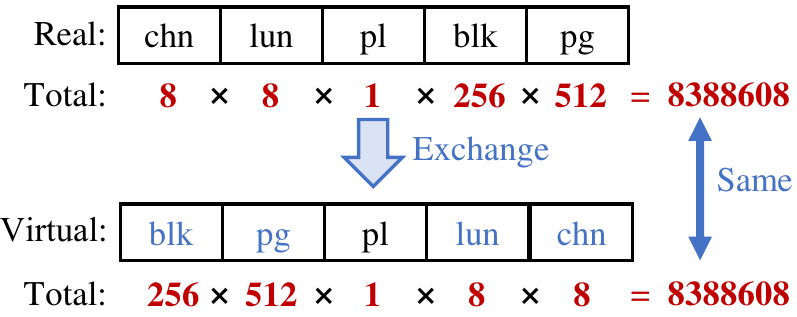}
    \caption{The principle of virtual PPN translation.}
    \label{fig-virtual}
\end{figure}

We propose virtual PPN representation to address the problem of non-contiguous PPNs caused by SSD internal parallelism. During model training of learned indexes, it is important to allocate contiguous PPNs for contiguous LPNs. However, the pages with consecutive LPNs may be written back to different flash chips, leading to non-contiguous PPNs.
To tackle this problem, LearnedFTL uses a VPPN representation to transform the non-contiguous PPNs scattered across different chips into contiguous ones. Figure~\ref{fig-virtual} shows the translation principle from PPN to virtual PPN. Since the total number of physical flash pages is fixed in an SSD, the PPN is formed in such a way that it represents the hierarchical tree structure of an SSD by the concatenation of address fields representing different levels of the hierarchy from the highest (channel) to the lowest (page) granularity. Because of the commutative law of multiplication, the order of these address fields in PPN can be changed to obey the allocation order. Thus, each physical page retains its unique number, and the new page number will become contiguous according to the allocation order.

\begin{figure}
    \centering
    \includegraphics[scale=0.85]{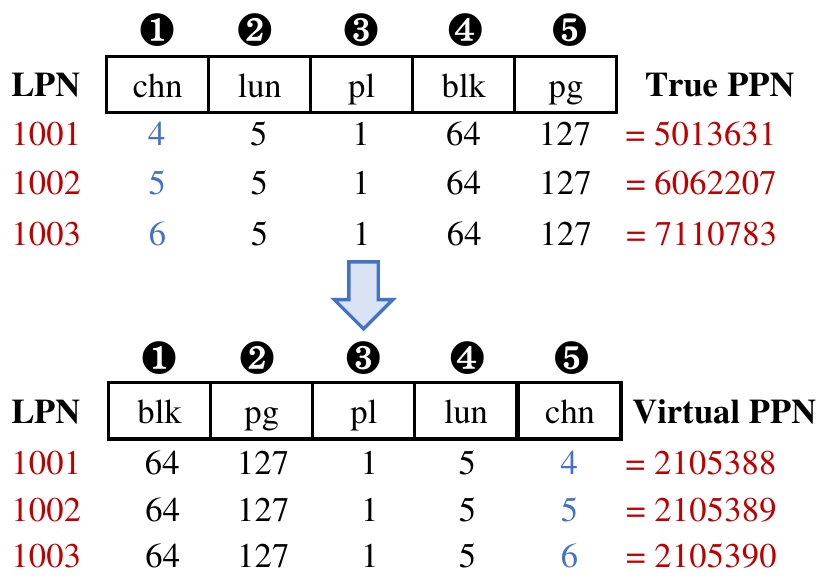}
    \caption{An example of PPN-to-VPPN translation.}
    \label{fig-virtual-example}
\end{figure}

Figure~\ref{fig-virtual-example} gives an example of the PPN-to-VPPN translation. In LearnedFTL, the allocation order is \emph{channel, chip, plane, page, and block}, which is the fastest allocation order based on the previous study~\cite{hu2011performance}. For requests with LPNs \emph{1001, 1002, 1003} that are already written to flash-based SSD, their PPNs are \emph{5013631, 6062207, 7110783}, which are not contiguous. However, after the PPN-to-VPPN translation by changing the order of the fields in the address appropriately, LearnedFTL obtains contiguous VPPNs \emph{2105388, 2015389, 2105390} for these LPNs.

The virtual PPN representation allows LearnedFTL to generate contiguous VPPNs for model training when valid pages are written to the flash-based SSDs concurrently. Since the training model is built based on LPN-VPPN mappings, the predicted VPPN needs to be translated back to PPN to obtain the physical flash page.

\subsection{Group-based Allocation Strategy}

Since random writes generate requests of non-contiguous LPNs, it's non-trivial to group them together and create a learned index during writes. Fortunately, garbage collection provides the opportunity to rearrange PPNs. Specifically, during GC, LearnedFTL can rearrange one GTD entry's PPNs to consecutive PPNs and then train models over these newly arranged PPNs.

However, the current dynamic allocation strategy used by LeaFTL and TPFTL makes PPN rearrangement difficult. This is because when allocating a flash page for a PPN, dynamic allocation will select the least busy flash chip to allocate pages for optimal parallelism and write efficiency. As a result, the PPNs of a GTD entry will be scattered across various locations. When building a learned model over this GTD entry via GC, LearnedFTL needs to collect the valid pages across multiple flash blocks, and these blocks also may contain PPNs belonging to other GTD entries. As a result, the GC process generates frequent data movement, which significantly increases the complexity and overhead of the model training process.

To address this PPN rearrangement issue, we propose a \textbf{group-based allocation strategy}. The basic idea is to divide GTD into groups of consecutive entries, referred to as \emph{GTD entry group}. Each group is allocated an exact number of contiguous flash blocks to accommodate all the LPNs of the group. When the flash blocks allocated to a GTD entry group are full, these used flash blocks are replaced by the same number of contiguous empty flash blocks. When there are no empty flash blocks or the cumulative number of flash blocks allocated to this GTD entry group reaches a threshold, GC is performed on the GTD entry group with the most invalid data pages. During GC, LearnedFTL reclaims data blocks by relocating the valid data pages and retrains the learned models for all GTD entries in this group.

\begin{figure}[t]
    \centering
    \includegraphics[scale=0.85]{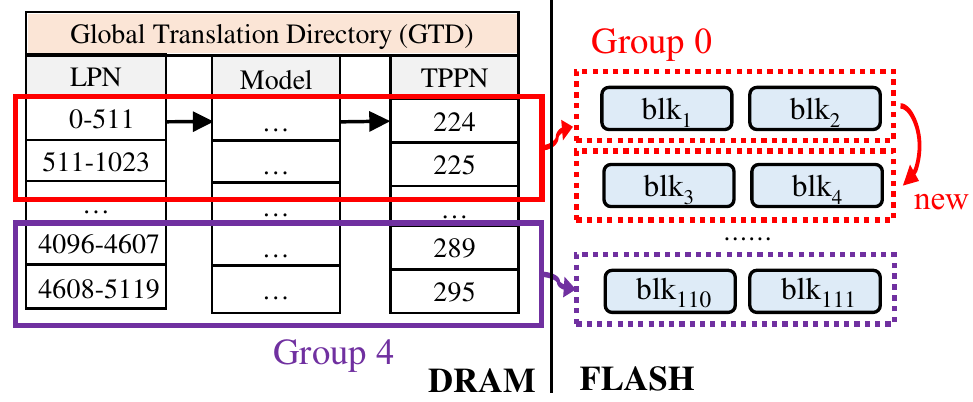}
    \caption{An example of group-based allocation.}
    \label{fig-group-allocation}
\end{figure}

Figure~\ref{fig-group-allocation} illustrates an example of group-based allocation. In this instance, for the convenience of presentation, each GTD entry group contains two entries and needs two contiguous flash blocks to accommodate all its LPNs. Therefore, LPNs \emph{0-1023} belong to group 0 and LPNs \emph{4096-5119} belong to group 4. When a request for data with LPN belonging to group 0 arrives, two contiguous blocks, \emph{blk1} and \emph{blk2}, are allocated to group 0. When a request for data with LPN belonging to group 4 arrives, another two contiguous blocks, \emph{blk110} and \emph{blk111}, are allocated to group 4 to accommodate the required data pages. When group 0 has no free physical pages, another two contiguous blocks, \emph{blk3} and \emph{blk4}, are allocated to this group. If group 0 is selected for garbage collection, all four blocks are collected directly.
 
A serious write-amplification concern arises with this group-based allocation strategy: when all GTD entry groups have been written at least once, a few hot GTD entry groups have been written frequently, which causes huge write amplification. To solve the problem, a global counter is associated with each GTD entry group to identify the hot groups by counting available free pages in the group.

To address the situation where hot GTD entry groups have limited or no free pages, LearnedFTL employs an \emph{opportunistic cross-group allocation} strategy. This approach allows these hot groups to utilize the available free-page spaces within flash blocks belonging to "cold" GTD entry groups that have an abundance of free pages and untrained models. By encroaching into the free-page spaces of the cold groups, LearnedFTL effectively avoids or delays the need for GC operations. Once the amount of encroachment reaches a specific threshold, GC is triggered for both the encroaching (hot) group and the encroached (cold) group. Subsequently, their respective models undergo retraining and training processes. Consequently, this opportunistic cross-group allocation approach not only reduces the frequency of GC and the write amplification caused by GC operations in hot groups but also ensures the early training of models in cold groups.

\subsection{Model Training}
\label{three-three}

To ensure the timeliness of the in-place-update model, LearnedFTL uses two model training strategies. One is sequential initialization, which is used to initialize the model through sequential write requests during data writing. The other is model training via GC, which is used to train the model during garbage collection to achieve higher accuracy.

\subsubsection{Sequential Initialization}
\label{sequential_init}

The main idea of sequential initialization is to update the learned index model in place based on sequential write requests. In many workloads, the I/O size of each request may range from several to tens of flash pages. When assigning contiguous PPNs for each I/O request, these LPN-PPN mappings can be seen as a \textbf{\emph{y=x}} model. Therefore, we can use these y=x models to update the corresponding in-place-update linear model. For each write request, there are four steps in sequential initialization:

\ding{172} \textbf{Obtaining contiguous PPNs.} LearnedFTL first writes the data of this request to the flash memory and obtains contiguous PPNs. After obtaining contiguous PPN, each LPN must check whether the corresponding bit in the bitmap is `1'. If it is, LearnedFTL updates it to `0'. 

\ding{173} \textbf{Generate the linear model.} LearnedFTL builds a \emph{\textbf{y=x}} model on these LPN-PPN mappings. Then LearnedFTL obtains the model's starting LPN ($\mathrm{LPN_{start}}$), ending LPN ($\mathrm{LPN_{end}}$), and length ($\mathrm{L}$). 

\ding{174} \textbf{Check corresponding model.} LearnedFTL locates the corresponding model with $\mathrm{<k, b, off>}$ in GTD by $\mathrm{LPN_{start}}$ and $\mathrm{LPN_{end}}$. After that, LearnedFTL calculates the length $\mathrm{L_{old}}$ of the existing model through the corresponding bitmap. 

\ding{175} \textbf{Update the model.} If $\mathrm{L_{old}}$ is smaller than $\mathrm{L}$, LearnedFTL performs in-place-update to replace the existing linear model with the newly generated linear model.

\subsubsection{Model Training via GC}
\label{model-training}

Since only long write requests will perform sequential initialization, LearnedFTL also proposes model training via GC to obtain a more comprehensive and accurate model.
With the help of group-based allocation, when LearnedFTL performs garbage collection, all valid pages of one GTD entry group can be collected and trained. 
When a GTD entry group needs to perform GC, the whole model training process via GC is divided into four steps:

\textbf{\ding {172} Regulate valid mappings.} First, LearnedFTL reads all the translation pages of this GTD entry group and only keeps the valid translations in memory. Then LearnedFTL sorts the valid translations by their LPNs to make them ordered. 

\textbf{\ding {173} Write valid pages back and obtain PPNs.} After regulating the valid LPNs, LearnedFTL allocates another group of flash blocks to this GTD entry group, then writes the valid pages back to the newly allocated flash blocks to get contiguous PPNs. 

\textbf{\ding {174} Train the learned model.} In this step, each GTD entry in this group will train its in-place-update linear model. For each GTD entry, calculate the offset of PPNs/LPNs from this GTD entry’s starting PPN/LPN. Then, perform greedy linear regression to fit to get the $<$k, b, off$>$ parameters array. 

\textbf{\ding {175} Evaluate the model.} After training the models, Learned FTL will evaluate the model and update the bitmap filter. During this process, each LPN will be inputted into the model. If the predicted PPN is accurate, the corresponding bit will be set to `1'.

\subsubsection{Model Training via Rewrite}

For some scenarios where GC rarely happens, the model training can be integrated into the SSD rewrite process~\cite{cai2015data,maneas2022operational}. The rewrite is a widely used reliability mechanism to reduce retention errors in modern SSDs by periodically reading, correcting, and reprogramming the flash memory. Rewrite happens frequently and is the most significant factor for write amplification~\cite{maneas2022operational}. During SSD rewrite, LPNs of flash pages can be sorted in order so that these pages are written back in contiguous PPNs, which then enables a model to be built and trained on them by LearnedFTL.

\subsection{Cost Analysis}

\label{three-cost}
Though LearnedFTL introduces multiple new components to apply the learned index in the FTL, it only introduces minor computational overhead, illustrated as follows.

\textbf{(1) Write}: For each write request, LearnedFTL incurs two additional operations, one is the \textbf{bitmap check} operation (Section~\ref{in-place-model}) to maintain the consistency of the model, the other is the sequential initialization (Section~\ref{sequential_init}). Both operations are performed in memory, and there are no calculation operations. Thus, the overhead can be ignored.

\textbf{(2) Read}: For each read request with an LPN, LearnedFTL incurs two additional operations when this LPN cannot hit in the CMT. The first operation is a \textbf{bitmap check} to check if this LPN can predict a real PPN. The second operation is a \textbf{model prediction} when the bitmap check is true. For these LPNs, LearnedFTL will use the model to predict the real PPN instead of an extra flash read. The model prediction includes calculating the VPPN with the \emph{y=kx+b} model and translating the predicted VPPN to PPN. 

\textbf{(3) GC}: The model training incurs two computational overheads during the GC period. The first one is \textbf{sorting} all the LPNs within each GTD entry (Step \ding {172} in Section~\ref{model-training}). The second one is \textbf{training} each GTD entry's model (Step \ding{174} in Section~\ref{model-training}).

Our experiments in Section~\ref{overhead-analysis} have detailed evaluations to quantitatively analyze these overheads.

\section{Performance Evaluation}
 \label{performance}

\subsection{Implementation and Experiment Setup}
\label{implementation}

\textbf{Experiment Setup}: The experiments are conducted on FEMU~\cite{li2018case}. FEMU is a QEMU-based and DRAM-backed SSD emulator that is widely used in recent studies~\cite{li2021loda, zhou2021remap, han2021zns+}. It runs in a machine with two Intel(R) Xeon(R) Gold 5318Y 2.10GHz CPUs and 128GB DRAM. The operating system is Linux with kernel version 5.4.0. The emulated SSD is configured with 32GB logical capacity plus 2GB over-provisioning space and has 64 parallel chips (8 channels and 8 ways per channel). Each flash chip has 256 flash blocks and each flash block has 512 flash pages. The size of a flash page is set to 4KB. The latency of NVMe SSD is 40 $\mu$s for NAND read, 200$\mu$s for NAND write, and 2ms for NAND erase, which are the default settings in FEMU and widely used in the recent flash-based studies~\cite{li2018case,li2021loda,han2021zns+}. Since the SSD rewrite for retention errors is not implemented in FEMU~\cite{maneas2022operational}, we only train models in GC.

LearnedFTL is compared against three representative page-level FTL designs, DFTL~\cite{gupta2009dftl}, LeaFTL~\cite{sun2023leaftl}, and TPFTL~\cite{zhou2015efficient}. We also use full-page mapping as a control (denoted as \emph{\textbf{ideal}}, which is considered a performance upper bound). In the experiments, we use both FIO benchmark~\cite{FIOmisc} and real-world applications/traces to evaluate different FTL designs.

\textbf{Prototype implementation}: We implement LearnedFTL by modifying the blackbox mode of the FEMU based on the TPFTL scheme. According to the allocation strategy and internal parallelism of the SSDs, we group each 64 consecutive GTD entries as a \emph{GTD entry group}. Since the size of a flash page is 4KB and each translation page has 512 LPN-PPN mappings, the GTD has 16384 entries. Each GTD entry group is allocated 64 flash blocks at a time, one for each of the 64 translation pages. For parameter setting in the piecewise linear model, 8 pieces are set by default. 

\begin{figure*}[t]
    \setlength{\abovecaptionskip}{0pt}
    \subfigure[Throughput]{
        \begin{minipage}[t]{0.35\linewidth}
            \centering
            \includegraphics[scale=0.76]{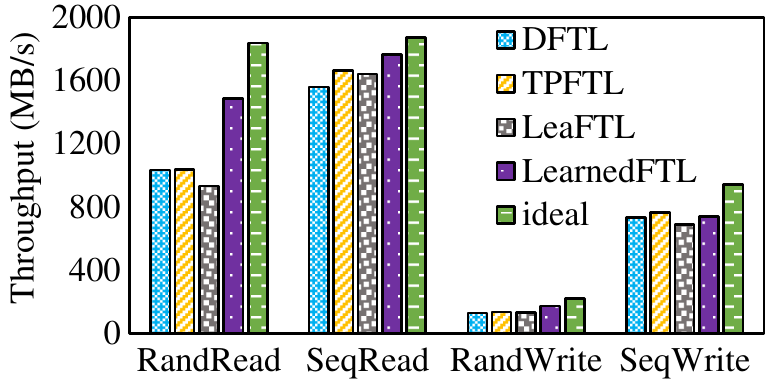}
            \label{fig-read-after-read}
        \end{minipage}%
    }
    \subfigure[Model and CMT hit ratio]{
        \begin{minipage}[t]{0.37\linewidth}
            \centering
            \includegraphics[scale=0.76]{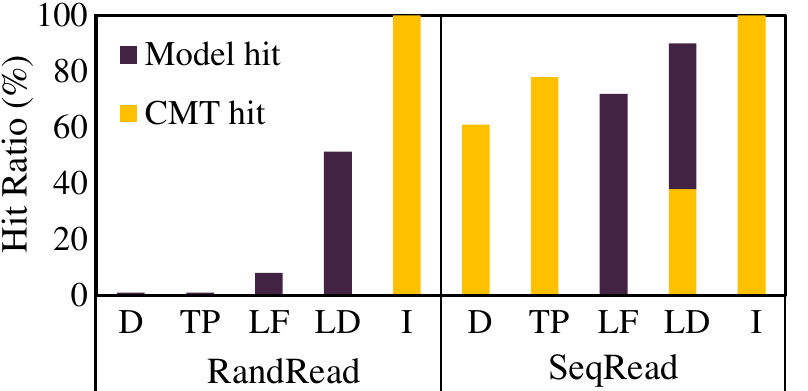}
            \label{fig-read-rand-hit}
        \end{minipage}
        
    }
    \subfigure[Write amplification]{
        \begin{minipage}[t]{0.23\linewidth}
            \centering
            \includegraphics[scale=0.76]{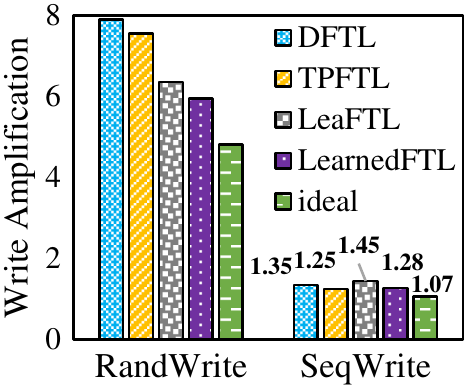}
            \label{fig-wa}
        \end{minipage}
    }    
    \caption{The FIO performance under 64 threads (D: DFTL, TP: TPFTL, LF: LeaFTL, LD: LearnedFTL, I: ideal FTL).}
    \label{double-read}
\end{figure*}

Since the previous demand-based FTLs, such as DFTL and TPFTL, and the recent LeaFTL are all implemented on trace-driven simulators, such as SSDsim~\cite{hu2011performance} and Flashsim~\cite{kim2009flashsim}, we incorporate them into the FEMU emulator according to their designs in the papers. We use FEMU's default greedy dynamic allocation strategy for their allocation strategy. Since LeaFTL's paper does not explain how the data in the model buffer is written to the SSD, in this article, the data writing strategy of LeaFTL will be consistent with that of TPFTL. Besides, we also added VPPN representation to LeaFTL to obtain continuous training data. We added and modified about 5,000 LoC to implement these baselines and the LearnedFTL in FEMU. The source code of these prototype implementations can be found in our Github repository~\cite{LearnedFTL}.

\textbf{Memory consumption}: Previous studies on demand-based FTLs usually set the capacity of CMT to about 3\% of the total number of page mappings~\cite{gupta2009dftl,zhou2015efficient}. For a fair comparison, we set the capacity of LeaFTL's model cache to have the same space overhead as the CMT of DFTL/TPFTL. For LearnedFTL, each model in the GTD entry has two parameters, \emph{$<$k,b,off$>$[N]} and \emph{bitmap}. For \emph{$<$k,b,off$>$[N]}, both \emph{k} and \emph{b} are set to a 2B float value (float16), and \emph{off} is set to a 2B integer value. For \emph{bitmap}, each slot is a bit, and there are 512 bits in total. To sum up, an in-place-update linear model requires 128 Bytes. After aggregating the space overhead of all the models, it can be calculated that the total overhead of the models in LearnedFTL is approximately half of the CMT space overhead in TPFTL and DFTL. Therefore, we set the CMT size of LearnedFTL to only accommodate 1.5\% of the total number of mappings to maintain the same memory overhead as other FTLs.

\begin{figure}[t]
    \centering
    \includegraphics[scale=0.9]{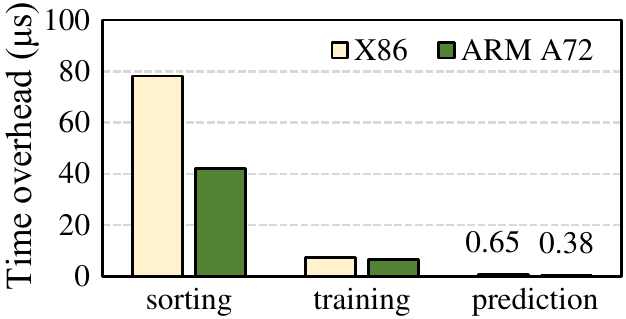}
    \caption{The computing overhead of the training operations between ARM and X86 processors.}
    \label{fig-compute-simulation}
\end{figure}

\textbf{Controller computing}: Since LearnedFTL adds some additional computing operations, it is necessary to correctly simulate the computing power of the SSD controller. The mainstream SSD controller CPUs are ARM's Cortex-A series and Cortex-R series. we compared the time consumption of executing the additional operations on the FEMU simulated CPU (X86) and a low-end embedded processor (ARM Cortex-A72), and each operation is at the maximum complexity. Figure~\ref{fig-compute-simulation} shows that the ARM A72 processor even performs better than the X86, which shows that we can use the X86 FEMU simulator to simulate LearnedFTL's computing power.

\subsection{FIO Benchmark}
\label{fio-sec}

We use the FIO benchmark~\cite{FIOmisc} to evaluate the performance of sequential writes, random writes, sequential reads, and random reads for different FTL designs. For each experiment, we ran at least three times to get the average results.

\emph{\textbf{(1) Read}}: 
For random-read and sequential-read evaluations, we first perform FIO random write and sequential write to warm up the whole SSD. Data is continuously written until the SSD is written over about 6 times to reach a stable state.  Since LeaFTL cannot handle 4KB random writes, the I/O size in the warm-up is 512KB (128 flash pages), which allowed the learned index of LeaFTL to be built normally. Then we perform a corresponding FIO read benchmark for evaluations. All the evaluations use 4KB I/O size and \emph{psync} I/O engine with 64 threads.

Figure~\ref{fig-read-after-read} illustrates the throughput results for d
ifferent FTL designs under different access patterns. For random read, LearnedFTL outperforms DFTL, TPFTL, and LeaFTL by 1.5$\times$, 1.4$\times$, and 1.6$\times$, respectively. For sequential read, LearnedFTL outperforms DFTL, TPFTL, and LeaFTL by 1.1$\times$, 1.1$\times$, and 1.1$\times$, respectively. Moreover, the performance of LearnedFTL is very close to that of the ideal FTL, achieving about 89.2\% and 96.8\% of the performance of the ideal FTL under random and sequential reads, respectively.

To explore the behind reasons, we also recorded the percentage of requests that hit the CMT and the learned models during random and sequential reads. The ideal FTL is used as a control which can be considered as an upper bound since its CMT has a hit ratio of 100\% and infinite space. For LeaFTL, we only count the situation that requires a single flash read (cache hit and model prediction is accurate), which is also marked as a model hit.

Figure~\ref{fig-read-rand-hit} shows that the CMT hit ratios of DFTL and TPFTL designs are almost 0 under random reads. The reason is that random reads show no locality, which makes the cache replacement policy fail to capture the access pattern. In LeaFTL, only 5\% of requests can perform a single flash read. Of the remaining requests, 43\% were triple reads and 52\% were double reads, so LeaFTL performs worst in random reads. By contrast, all learned index models in LearnedFTL can be stored in SSD's memory with 55.5\% accuracy, which means 55.5\% extra flash translation reads can be reduced. Thus, it can significantly improve the random-read performance over other FTL schemes.

Under sequential reads, LearnedFTL still outperforms DFTL, LeaFTL, and TPFTL. Since each thread competes for cache space, DFTL, TPFTL, and LeaFTL only achieve 61\%, 80\%, and 76\% hit ratios on CMT and models. By contrast, because all models of LearnedFTL can be stored in SSD memory, LearnedFTL can resolve contentions effectively. LPN misses in CMT can be hit in the model, and no cache replacement will occur. As a result, LearnedFTL achieves a combined CMT-Model hit ratio of up to 90\%, eliminating 90\% of the LPN-PPN double reads. Thus, LearnedFTL achieves the best performance among all FTLs and approaches that of the ideal FTL, which is the upper bound.

\begin{figure}[t]
    \centering
    \includegraphics[scale=0.85]{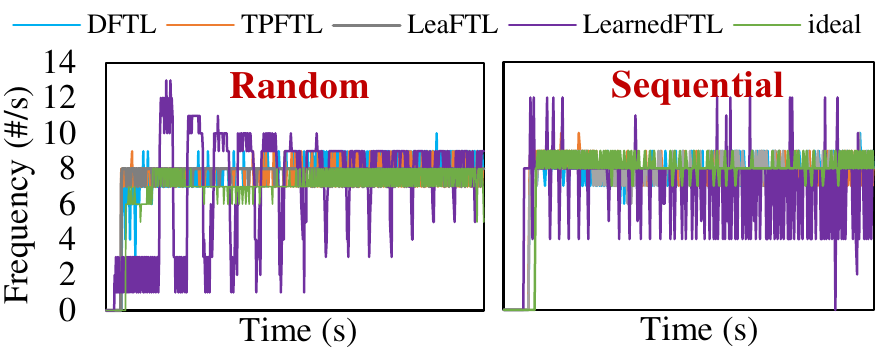}
    \caption{The GC frequency of all FTL designs under FIO random and sequential write benchmarks.}
    \label{fig-gc-frequency}
\end{figure}

\emph{\textbf{(2) Write}}:
We perform FIO write for the random-write and sequential-write evaluations, and all the evaluations use 4KB I/O size and \emph{psync} I/O engine with 64 threads. 

Figure~\ref{fig-read-after-read} shows that under random writes, LearnedFTL outperforms other schemes by 1.2$\times$ to 1.4$\times$, respectively, because of LearnedFTL's group-based allocation strategy. With the group-based allocation strategy, LearnedFTL selects one GTD entry group for each GC, only the translation pages of this GTD entry group need to be updated. That is, a maximum of 64 translation pages are updated per GC. However, for the dynamic allocation used in other schemes, when the same number of data blocks are collected, the LPN range of flash pages written back may be more than 64 translation pages, incurring additional write amplification. 

Owing to the spatial locality of sequential writes, Learned FTL performs almost the same as DFTL and TPFTL, by less than 2\%. Unlike the dynamic allocation strategy which selects the blocks with the fewest valid pages in each GC, the group-based allocation strategy performs GC on a group-by-group basis, which may result in more valid pages being written back. 
Fortunately, the opportunistic cross-group allocation allows the hot GTD entry group in sequential writes to use free pages of cold GTD entry groups, reducing the number of valid pages being written back. Thus, LearnedFTL's sequential write performance is the same as other FTLs.

\begin{figure}[t]
    \centering
    \includegraphics[scale=0.85]{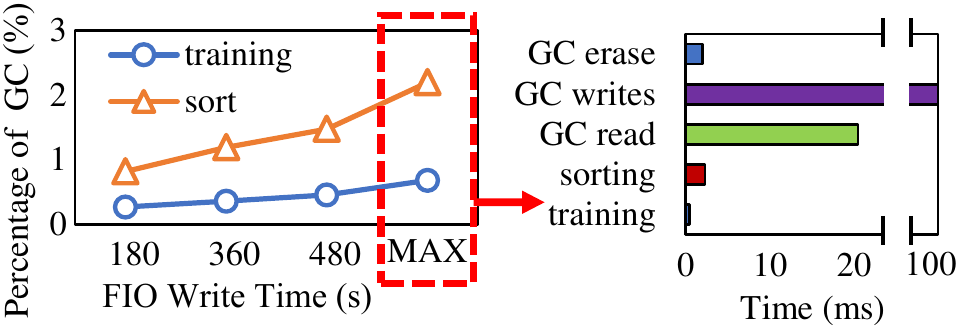}
    \caption{The time overhead of sorting and training under different running times of FIO random writes (MAX means almost all pages are valid during GC).}
    \label{fig-gc-overhead}
\end{figure}

\subsection{Overhead Analysis}

\label{overhead-analysis}
We evaluate the overhead induced by additional operations that LearnedFTL brings as mentioned in Section~\ref{three-cost}.

\textbf{(1) GC frequency and write amplification}: In LearnedFTL, model training happens in GC, and LearnedFTL proposes group-based allocation to assist model training. Therefore, the GC frequency and write amplification are critical indicators. Figure~\ref{fig-gc-frequency} illustrates the GC frequency of various FTLs in the FIO write evaluations. Although the GC frequency of LearnedFTL fluctuates, the total number of GCs triggered under random writes and sequential writes of LearnedFTL (4188 and 4285) are less than DFTL (4335 and 4572), LeaFTL(4395 and 4473) and TPFTL (4335 and 4304).
Figure~\ref{fig-wa} shows that the write amplifications of DFTL and LeaFTL are larger than LearneFTL in random writes because the group-based allocation requires fewer translation page writes. For sequential writes, with the assistance of opportunistic cross-group allocation, the write amplification of LearnedFTL is comparable to other FTLs.
In summary, our group-based allocation can effectively assist the model training without inducing additional GC and write amplifications.

\textbf{(2) Overhead of training and sorting}: The \textbf{model training} (denoted as training) and \textbf{LPNs-sorting} (denoted as sorting) are two additional operations added to GC in LearnedFTL. In our implementation, we group 64 GTD entries into one group. During each GC, a maximum of 64 LPNs-sorting and model training operations will be triggered for each GTD entry group. Figure~\ref{fig-compute-simulation} shows each GTD entry needs about 50$\mu$s for sorting and training in ARM Cortex-A72. The maximum additional overhead incurred by sorting and training is equivalent to about 80 SSD reads (40 $\mu$s per read), which is negligible since GC for each GTD entry group will incur tens of thousands of SSD reads and writes. Figure~\ref{fig-gc-overhead} shows that the time overhead of sorting and training only accounts for up to 3.2\% of the GC execution time.

\begin{figure}[t]
    \subfigure[Write and GC]{
        \begin{minipage}[t]{0.5\linewidth}
            \centering
            \includegraphics[scale=0.8]{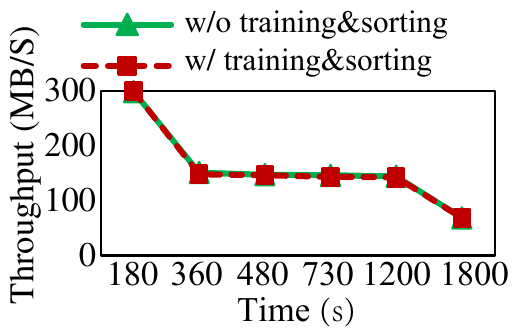}
            \label{fig-write-overhead}
        \end{minipage}%
    }
    \subfigure[Read]{
        \begin{minipage}[t]{0.4\linewidth}
            \centering
            \includegraphics[scale=0.8]{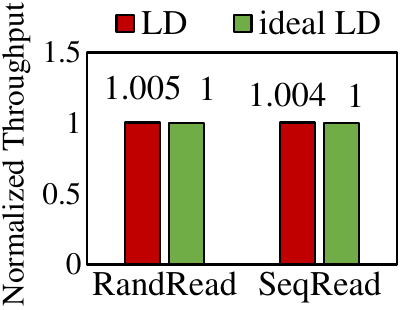}
            \label{fig-read-overhead}
        \end{minipage}
    }    
    \caption{LearnedFTL with and without additional computing operations (LD: LearnedFTL, ideal LD: ideal LearnedFTL.}
    \label{overhead-two}
\end{figure}

To further explore whether they will introduce additional latency, we compare the FIO random write performance of LearnedFTL with and without training and sorting operations. Figure~\ref{fig-write-overhead} shows that their performance difference is nearly negligible (less than 0.7\%), further verifying that the computing overhead of training and sorting is minimal in LearnedFTL.

\textbf{(3) Overhead in read operations}: Only LPNs that can be correctly predicted will perform \textbf{model prediction} (0.65$\mu$s in Figure~\ref{fig-compute-simulation}). This means there is no miss penalty in model predictions. Although there is no miss penalty, if the model prediction takes too long, it will reduce the advantage of reducing double reads. We implement the ideal LearnedFTL, which puts all mappings in memory. For ideal LearnedFTL, each time the bitmap check is yes, it can directly get the PPN through the mapping table without model prediction. Figure~\ref{fig-read-overhead} shows that the FIO read performance gap between LearnedFTL and ideal LearnedFTL does not exceed 1\%, demonstrating that the model predictions are lightweight.

\subsection{Real-World Applications}

\begin{figure}[t]
    \subfigure[Normalized throughput]{
        \begin{minipage}[t]{0.45\linewidth}
            \centering
            \includegraphics[scale=0.77]{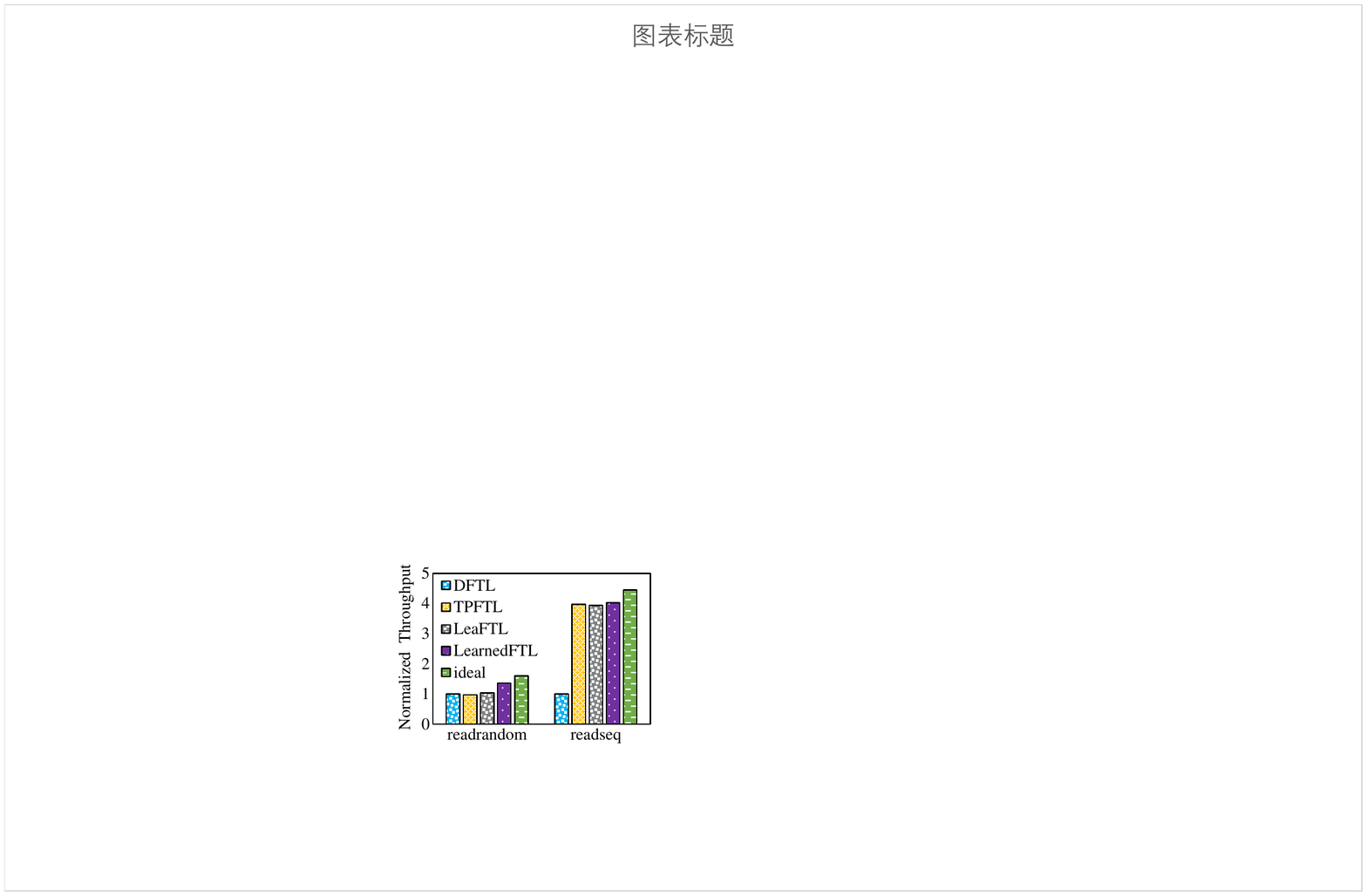}
            \label{rocksdb_throughput}
        \end{minipage}%
    }
    \subfigure[CMT and model hit ratio]{
        \begin{minipage}[t]{0.46\linewidth}
            \centering
            \includegraphics[scale=0.75]{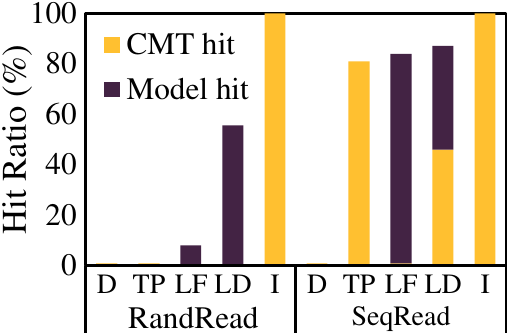}
            \label{rocksdb_hit_ratio}
        \end{minipage}
    }
    \caption{RocksDB performance with one thread (D: DFTL, TP: TPFTL, LF: LeaFTL LD: LearnedFTL, I: ideal FTL). }    
    \label{rocksdb_performance}
\end{figure}

\textbf{RocksDB}: RocksDB\cite{RocksDB} is a widely used LSM-Tree-based KV store designed to exploit the parallelism of flash-based SSDs. As we mentioned before, LSM-Trees can merge random writes into sequential ones, but at the cost of relatively poor services to random reads. We deploy RocksDB with EXT4 file system on top of each FTL design and use the \emph{db\_bench} tool of RocksDB with one thread, which is consistent with the previous studies~\cite{ kannan2018redesigning, raju2017pebblesdb}. To evaluate the read performance, we first use the \emph{fillseq} and \emph{overwrite} in $db\_bench$ to write the DB to 80\% full, then we perform \emph{readrandom} and \emph{readseq} in $db\_bench$ to evaluate the read performance in RocksDB. 

In terms of throughput, Figure~\ref{rocksdb_throughput} shows that LearnedFTL outperforms other FTLs by 1.3$\times$ $\sim$ 1.4$\times$ in random reads. LearnedFTL also outperforms other FTLs by 1.02$\times$ $\sim$ 4.0$\times$ in sequential reads. To better understand these results, Figure~\ref{rocksdb_hit_ratio} shows the model and CMT hit ratios recorded in these evaluations. In a single-threaded environment, DFTL does not exploit and thus fails to benefit from the spatial locality, so its CMT hit ratio is zero. TPFTL and LeaFTL can achieve an 81\% CMT hit ratio and 83\% model hit ratio by exploiting the spatial locality. By contrast, since LearnedFTL exploits both the spatial locality and the learned model, it achieves 0.3\% and 46\% CMT hit ratio, 55\% and 41\% model hit ratio in random reads and sequential reads, respectively. 

\begin{table}[t] 
\small
\begin{center}   
\caption{Filebench configurations.} 
\label{table_filebench_configuration} 
\begin{tabular}{|c|c|c|c|}   
\hline   \textbf{Name} & \textbf{Fileset} & \textbf{Feature} & \textbf{Threads} \\   
\hline   fileserver & 225,000 $\times$ 128KB & write heavy & 50 \\
\hline   webserver & 825,000 $\times$ 16KB & read heavy & 64 \\ 
\hline   varmail & 475,000 $\times$ 16KB & all read & 64  \\
\hline
\end{tabular}   
\end{center}
\end{table}

\begin{figure}[t]
\centering
\includegraphics[scale=0.9]{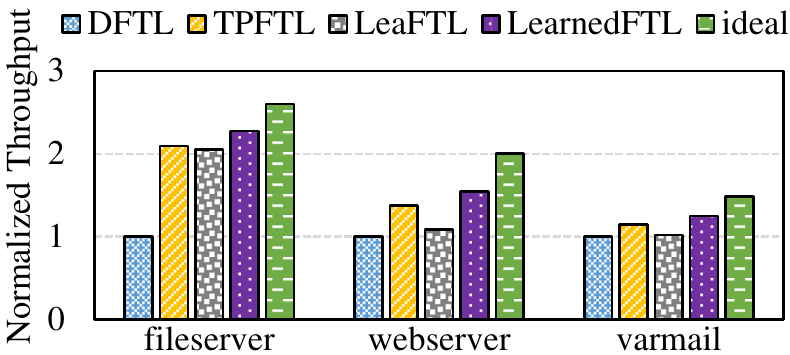}
\caption{The normalized throughput of Filebench.}
\label{fig-filebench}
\end{figure}

\textbf{Filebench}: Filebench~\cite{Filebench} is a highly flexible storage benchmark. We select three workloads that are most widely used in previous studies~\cite{zhou2021remap, han2021zns+, bjorling2021zns}: 
\emph{fileserver} (write heavy), \emph{webserver} (read heavy, less random write), and \emph{varmail} (read:write=1:1). Their configurations, consistent with previous studies~\cite{han2021zns+, zhou2021remap}, are summarized in Table~\ref{table_filebench_configuration}.

Figure~\ref{fig-filebench} shows the normalized performance of the four FTLs. LearnedFTL outperforms other schemes by 1.1$\times$ to 2.3$\times$. As we mentioned in Challenge \#1 in Section~\ref{two-four}, the inaccuracy of LeaFTL's learned models makes LeaFTL still require many double reads under these workloads with high locality. As a result, LeaFTL's performance is lower than TPFTL and LearnedFTL. Since LearnedFTL preserves the CMT, so most requests with high locality can be hit directly through the CMT. In addition, the learned index models can also make predictions for requests that cannot be hit in the CMT, further improving performance.

\subsection{Real-world Traces}

We select four traces (three WebSearch traces and one Systor trace) to evaluate the efficacy of different FTL designs. The three WebSearch traces are read-intensive workloads that are generated from a popular search engine~\cite{Oltp}. The Systor trace is the enterprise storage traffic on modern commercial office VDI for 28 days~\cite{Lee2017Understanding}. The four traces all have strong locality. For these traces, we pick the busiest periods (20 minutes to 2 hours). Since the WebSearch traces are relatively old, we scale up them to reflect modern SSD workloads~\cite{li2021loda}. The workload characteristics of the four traces are summarized in Table~\ref{table_trace_configuration}. Before replaying the four traces, we warm up the whole SSD to a steady state with the same warm-up method mentioned in Section~\ref{fio-sec}. We choose TPFTL and LeaFTL as the baselines for the tail latency evaluation.

\begin{table}[t]   
\small
\begin{center}   
\caption{Workload characteristics of four traces.}
\label{table_trace_configuration} 
\begin{tabular}{|c|c|c|c|}   
\hline   \textbf{Traces} & \textbf{\# of I/O} & \textbf{Average I/O size} & \textbf{Read ratio} \\   
\hline   WebSearch1 & 1,055,235 & 15.5KB & 100\%  \\
\hline   Websearch2 & 1,200,964 & 15.3KB & 99.98\%  \\ 
\hline   Websearch3 & 793,073 & 15.7KB & 99.96\%  \\
\hline   Systor17 & 1,253,423 & 10.25KB & 61.6\% \\
\hline
\end{tabular}   
\end{center}  
\end{table}

\begin{figure}[t]
\centering
\includegraphics[scale=0.9]{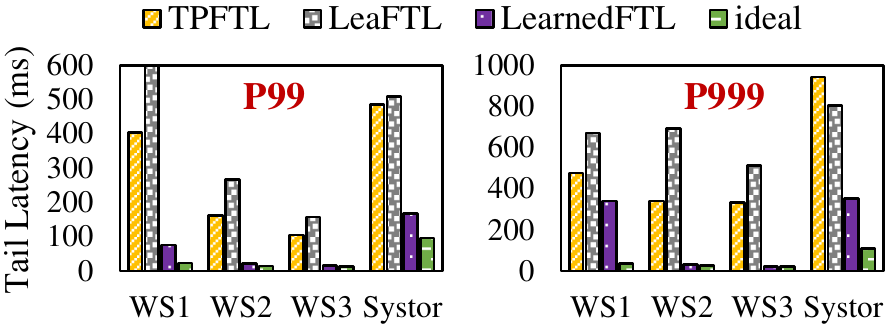}
\caption{The P99 and P999 tail latencies results under four traces (WS\# denotes WebSearch\#).}
\label{fig-tail}
\end{figure}

\begin{figure}[t]
\centering
\includegraphics[scale=0.76]{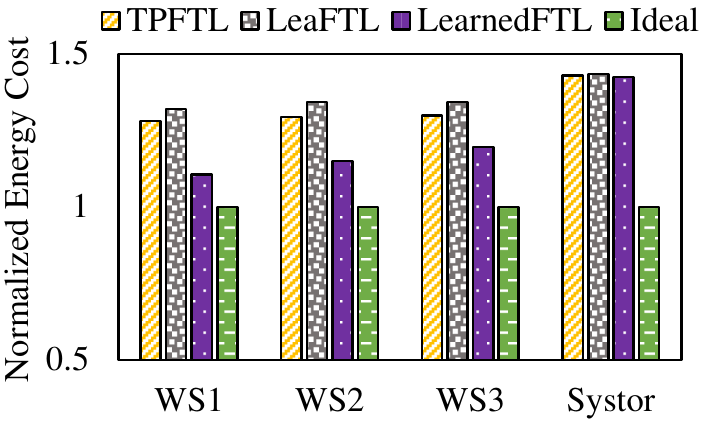}
\caption{The energy cost under four traces (WS\# denotes WebSearch\#).}
\label{energy-cost}
\end{figure}

Figure~\ref{fig-tail} shows the \emph{P99} and \emph{P99.9} tail latencies of TPFTL, LearnedFTL, and ideal FTL driven by the four traces. Under these four traces, compared to TPFTL, LearnedFTL reduces the \emph{P99} tail latency by 5.3$\times$, 7.4$\times$, 6.5$\times$, and 2.9$\times$, respectively, with an average of 5.5$\times$. Compared with LeaFTL, LearnedFTL reduces the \emph{P99} tail latency by 7.8$\times$, 12.2$\times$, 9.7$\times$, and 3.0$\times$, respectively, with an average of 8.2$\times$. Moreover, compared with TPFTL and LeaFTL, LearnedFT reduces the \emph{P99.9} tail latency by up to 13.9$\times$ and 21.4$\times$, respectively. LearnedFTL’s tail latency in WebSearch2 and WebSearch3 are extremely close to that of the ideal FTL. Although TPFTL and LeaFTL can maintain high CMT hit ratios or model hit ratios on workloads with strong locality, sporadic double reads and triple reads still induce high tail latency. By contrast, LearnedFTL's learned model can further reduce these sporadic double reads by accurate PPN prediction, thus reducing tail latency.

\subsection{Energy Cost}
We established a basic power/energy model based on NANDFlashSim\cite{NANDFlashSim-TOS16} and conducted tests. Figure~\ref{energy-cost} provides a comparison of energy consumption for each FTL in four real traces. In the three WebSearch traces, LearnedFTL reduces energy consumption by 1.09$\times$ to 1.2$\times$ than TPFTL and LeaFTL, respectively. They perform similar under Systor trace. The reason is that the energy consumption of flash write and erase overwhelms that of flash read. In workloads that are not read-intensive, the reduction in reading energy consumption has limited impact on the total energy consumption. In read-intensive workloads, LearnedFTL can reduce energy consumption compared to other FTLs.

\section{Discussion}
\label{discussion}
\textbf{Linearity in page-level mappings}. 
One common concern is why simple linear models can effectively fit page-level mappings. In real-world workloads, write requests often consist of multiple consecutive LPNs. When these LPNs are written to consecutive PPNs, we can observe a linear relationship like \emph{y=x+b}. Furthermore, GC can also contribute to the linearity. During the GC process, the FTL can collect LPNs, sort them, and write them back to consecutive physical pages, resulting in a linear relationship such as \emph{y=kx+b}. The existence of this linearity enables good fitting results of simple linear models. Moreover, these models are easy to train and require less training time. Consequently, employing linear models for fitting purposes emerges as the optimal choice.

\textbf{Efficiency in random access}. The superior performance of LearnedFTL in random access comes from its unique learning pattern. Unlike traditional machine learning methods that try to fit access patterns~\cite{yoo2020reinforcement,zhang2019reinforcement}, the learned index models in LearnedFTL fit the relationship between LPN-PPN, that is, the relationship between data and actual locations. In this way, whether it is random access or regular access, the model can stably calculate the PPN corresponding to the LPN, thereby significantly improving the performance of random access.

\textbf{Model's Space Overhead}. The space overhead of learned index models is an additional consideration introduced in FTL. Therefore, how to further reduce the space consumption of the model is a future work. Since LearnedFTL uses rounding mode to calculate PPN and bitmap filter to ensure the accuracy of predictions, the computational precision requirement is not high. As a result, it is possible to consider using lower-precision data types, such as Float8, for model parameters. Apart from that, we can further reduce the space overhead by compressing each model' bitmap through compression or encoding techniques.

\section{Conclusion}
\label{conclusion}

We propose LearnedFTL, a learning-based page-level FTL design, by exploiting some unique characteristics of both the flash device and the learned index. LearnedFTL uses an in-place-update linear model to build learned indexes efficiently, a virtual PPN representation to obtain contiguous PPNs for sorted LPNs, and a group-based allocation and model training via GC/rewrite strategy to reduce the training overhead. Our FEMU-based prototype and extensive evaluations have validated that LearnedFTL outperforms the state-of-the-art TPFTL and LeaFTL schemes.

\section*{Acknowledgements}
We thank the anonymous reviewers and You Zhou from Huazhong University of Science and Technology for their valuable feedback. This work is supported by the National Key Research and Development Program of China (Grant No. 2023YFB4502703), the National Natural Science Foundation of China under Grant No. U22A2027 and No. 61972325, and Open Research Projects of Zhejiang Lab (No. 2021DA0AM01/002). Dr. Zhang is supported in part by the National Natural Science Foundation of China under Grant No. 62332021 and the Fundamental Research Funds for the Central Universities, Peking University.

\bibliographystyle{IEEEtranS}
\bibliography{refs}

\end{document}